\documentclass[11pt]{JHEP3}
\date{}

\usepackage{graphicx}

\usepackage{amsfonts}
\usepackage{epsfig}
\newcommand{\be}{\begin{eqnarray}}
\newcommand{\ee}{\end{eqnarray}}
\newcommand{\bi}{\begin{itemize}}
\newcommand{\ei}{\end{itemize}}

\title{Fast Electroweak Symmetry Breaking\\ and Cold Electroweak Baryogenesis}

\author{Kari Enqvist\\Physics Department and Helsinki Institute of Physics\\FI-00014 University of Helsinki, Finland\\ E-mail: \email{kari.enqvist@helsinki.fi}}
\author{Philip Stephens\\ Physics Department, Lancaster University\\ Lancaster LA1 4YB, United Kingdom\\ E-mail: \email{p.stephens@lancaster.ac.uk}}
\author{Olli Taanila\\ Helsinki Institute of Physics\\FI-00014 University of Helsinki, Finland\\ E-mail: \email{olli.taanila@iki.fi}}
\author{Anders Tranberg\\ Helsinki Institute of Physics\\FI-00014 University of Helsinki, Finland\\ and Department of Physical Sciences\\FI-90014 University of Oulu, Finland\\ E-mail: \email{anders.tranberg@helsinki.fi} }

\abstract{We construct a model for delayed electroweak symmetry breaking that takes place in a cold Universe with $T\ll 100\,$GeV
and which proceeds by a fast quench rather than by a conventional, slow, phase transition. This is achieved by coupling the Standard Model Higgs to an additional scalar field.
We show that the quench transition can be made fast enough for successful Cold Electroweak Baryogenesis, while leaving known electroweak physics unchanged. }

\keywords{Electroweak symmetry breaking, Baryogenesis, Tachyonic preheating}
\preprint{HIP-2010-13/TH}

\begin{document}


\section{\label{sec:theintroduction}Introduction}

It is well known that electroweak baryogenesis within the Standard Model is subject to a double kill:
the electroweak phase transition is a cross-over \cite{Kajantie:1996mn} so that the requirement for non-equilibrium is
never fulfilled; and at the electroweak phase transition temperatures, the CP-violation in the fermion mass matrix
is too small by many orders of magnitude (see \cite{Rubakov:1996vz} for a review).
However, it has also been known for some time that the observed baryon asymmetry could be explained
if the electroweak transition is cold rather than hot \cite{GarciaBellido:1999sv,Krauss:1999ng,Copeland:2001qw,Tranberg:2003gi}
with $T\ll 100\,$GeV, and if the transition
itself is a fast quench rather than adiabatic \cite{Tranberg:2006ip,Tranberg:2006dg}. Moreover,
recent analytical work \cite{Hernandez:2008db} and numerical simulations have indicated that in such a cold environment, Standard Model CP-violation
is much larger than at electroweak temperatures and is indeed strong enough
to account for the baryon asymmetry \cite{Tranberg:2009de}. In Cold Electroweak Baryogenesis, most of the baryon asymmetry is produced at the initial quench when the Higgs field 
is rapidly falling down the slope of the potential. Baryon production essentially stops after the first oscillation,
after which the coherent Higgs field starts decaying, thereby reheating the Universe. The reheating temperature should be low enough that sphaleron diffusion does not subsequently wipe out the baryon number.

In the early Universe the rate of any global change is necessarily related to the Hubble rate,
which at the electroweak scale is very small. Hence, in the absence of a first order phase transition, the realization of a fast quench does not appear to be possible within the Standard Model. But if the Higgs field were coupled to some beyond-the-Standard Model fields, the situation might change. Indeed, the Higgs could be just one of
many scalar fields such as in the low energy limit of string theory. 

The inflaton field could also be an example of such ``moduli'',
and a low temperature state and a fast quench
could both follow from single field ("inverted hybrid"-type) low-scale inflation \cite{Knox:1992iy,German:2001tz,Copeland:2001qw,vanTent:2004rc}.
However, this is at the expense of allowing a rather general inflaton potential (up to the sixth power in the field), as well as some
fine tuning of parameters. Presumably this tuning can be alleviated by considering more complicated models (see, for instance \cite{Ross:2010fg}).

Alternatively, as in the present paper, one can assume that inflation is decoupled from the electroweak phase transition and takes place at some
high energy scale. We then study the generic conditions for the cold, quenched electroweak phase transition in the
presence of an extra scalar field coupled to the Standard Model Higgs. In particular, we focus on
the possibility of triggering the quench through the cosmological expansion,
via thermal corrections to the effective potential. We will demonstrate both that, under certain conditions,
the phase transition can be delayed
until $T\simeq {\cal O}(1)\,$GeV, and that the transition can be fast (in a sense to be specified below).

We will describe briefly the problem of the fast quench in the next section. In section \ref{sec:triggerhiggs}
we couple the Higgs field to another scalar field and show that a fast quench can be achieved.
In section \ref{sec:triggersigma} we show how to trigger a quench of the $\sigma$ through finite temperature corrections to the mass and the expansion of the Universe. We then briefly consider effects such as defect formation, bubble nucleation and loop corrections in section \ref{sec:checks}. In section \ref{sec:electroweak} we introduce the criteria to be fulfilled in order that the Standard Model physics remains unchanged and section \ref{sec:theconclusion} contains our conclusions.


\section{\label{sec:theproblem}The Higgs quench in the Standard Model}

Consider the standard cosmological scenario, with inflation leading to reheating of the Standard Model degrees of
freedom, and a reheating temperature $T_{\rm reh}\gg 100\,$GeV.
The Higgs potential is (neglecting for the moment interaction with fermions and gauge fields),
\be
V(\phi)=V_0-\frac{\mu^2}{2}\phi^2+\frac{\lambda}{4}\phi^4.
\ee
At zero temperature, the symmetry is "broken", in the sense that the field acquires a vacuum expectation value $
v=\mu/\sqrt{\lambda}$,
which in the Standard Model has the value $v=246\,$GeV. At finite temperature, the symmetry
is "restored" through a thermal mass (at leading order in coupling constants), so that
\be
v^2(T)=\frac{\mu^2-m_{\rm th}^2(T)}{\lambda}, \qquad m^2_{\rm th}(T)\simeq\kappa\, T^2,
\label{eq:Tvev}
\ee
where $\kappa<1$ is a constant\footnote{This is the $T\gg m$ result, which we use for illustration. A more refined treatment would include the effects of a finite mass (see also section \ref{sec:triggersigma}).},
assumed to include interactions with all the Standard Model degrees of freedom. When the right-hand side
of (\ref{eq:Tvev}) is negative, the expectation value is zero. Since the temperature decreases with the
scale factor $a$ as $T\propto 1/a$, we have that the speed of the quench can be quantified as
\be
v_q=\frac{1}{2\mu^3}\frac{d}{dt}\left(m_{\rm th}^2(T)-\mu^2\right)_{T=T_q}=\frac{H_q}{\mu},
\ee
where the subscript $q$ denotes the time of the quench, $m^2_{\rm th}(T_q)=\mu^2$.
At the electroweak scale $V_0^{1/4}\simeq \mu \simeq T_q\simeq 100\,$GeV while the Hubble rate is very small so that
the quench rate is minuscule with $v_q=H_q/\mu \simeq 10^{-16}$.
Numerical simulations suggest that for Cold Electroweak Baryogenesis to reproduce the observed baryon asymmetry, we must require \cite{Tranberg:2006dg}
\be
v_q>0.1, \qquad T_q\simeq 1\,\textrm{GeV}.
\ee
Clearly, in the Minimal Standard Model this scheme does not work, and the electroweak transition is an equilibrium cross-over. Note also that even ignoring the
constraint on $v_q$, $T_q\simeq 1\,$GeV would require the thermal correction strength
$\kappa\simeq 10^4$, since the Higgs mass $m_H^2=2\mu^2\simeq (100-200\,$GeV$)^{2}$ is fixed.


\section{\label{sec:triggerhiggs}Quench in the presence of an additional scalar}

Consider now a system of two scalar fields; the Higgs, $\phi$, and a Standard Model singlet field, $\sigma$. The potential is chosen to be
\be
\label{eq:potential}
V(\sigma,\phi)=V_0-\frac{\lambda_4}{4}\sigma^4+\frac{\lambda_6}{6}\sigma^6-\frac{g^2}{2}\sigma^2\phi^2+\frac{m_\phi^2}{2}\phi^2+\frac{\lambda_\phi}{4}\phi^4.
\ee
Note that the potential is not renormalizable ($\sigma^6$), as it turns out that a renormalizable potential ($\sigma^2$, $\sigma^4$) does not lead to a fast enough quench (see also \cite{Copeland:2001qw,vanTent:2004rc}). Denoting by $(v_\sigma,v_\phi)$ the global minimum, we have the constraints
\be
\label{eq:vevs}
-g^2v_\phi^2-\lambda_4 v_\sigma^2+\lambda_6v_\sigma^4=0,\\
m_\phi^2-g^2v_\sigma^2+\lambda_\phi v_\phi^2=0,
\ee
and to avoid a spurious cosmological constant, we impose that in the minimum, the potential should vanish,
\be
V(v_\sigma,v_\phi)=0.
\ee
This means that
\be
g^2v_\sigma^2&=&m_\phi^2+\lambda_\phi v_\phi^2,\\
\lambda_4v_\sigma^4&=&12V_0+2m_\phi^2v_\phi^2-\lambda_\phi v_\phi^4,\\
\lambda_6v_\sigma^6&=&12V_0+3m_\phi^2v_\phi^2.
\label{eq:constraint123}
\ee
We fix $v_\phi=246\,$GeV (electroweak physics), $V_0=100^4\,$GeV$^4$ (to end up in the broken electroweak phase after thermalisation) and $\lambda_\phi=m_H^2/(2v_\phi^2)$, $m_H=160\,$GeV (Higgs mass allowed by experiment). This leaves the free parameters $v_\sigma$ and $m_\phi$. 

Imagine now starting off at $\phi=0$, $\sigma=+\epsilon\ll 1$. As long as $\sigma<\sigma_q=m_\phi/g$, $\phi=0$ is enforced. Then upon reaching $\sigma_q$, the effective mass of $\phi$ flips sign, and $\phi$ will go through a spinodal transition and electroweak symmetry breaking. We are interested in the speed of the quench, and so we calculate
\be
v_q=\frac{1}{2\mu^3}\frac{d}{dt}\left(m_\phi^2-g^2\sigma^2\right)_{\sigma=m_\phi/g},
\ee
where $\mu=100\,$GeV is just a normalisation scale, in order to compare with \cite{Tranberg:2006dg}. For effective baryogenesis, we require $v_q>0.1$. We easily find
\be
v_q=\frac{m_\phi^2}{\mu^2}\frac{\dot{\sigma}}{\mu \sigma}.
\ee
Simple energy considerations show that at $\sigma_q$, the velocity of $\sigma$ is
\be
\dot{\sigma}\simeq\sqrt{\frac{2\lambda_4}{4}\left(\frac{m_\phi}{g}\right)^4-\frac{2\lambda_6}{6}\left(\frac{m_\phi}{g}\right)^6}.
\label{eq:sigmadot}
\ee
Fig.~\ref{fig:vq} shows $v_q$ in $v_\sigma-m_\phi$ space. Only the range $v_q>0.1$ is included. Let us for the purpose of illustration choose a representative point, $v_\sigma=3000\,$GeV, $m_\phi=150\,$GeV (see also Fig.~\ref{fig:final}). In that case, we have
\be
v_\sigma=3000\textrm{ GeV},&\quad& m_\phi=150\textrm{ GeV}\quad\rightarrow\quad\nonumber\\
 v_q=0.127,&&\lambda_4=3.89\times 10^{-5},\quad \lambda_6=7.25\times 10^{-12}\textrm{ GeV}^{-2},\quad g^2=3.92\times 10^{-3}.\nonumber\\
\ee
\EPSFIGURE[t]{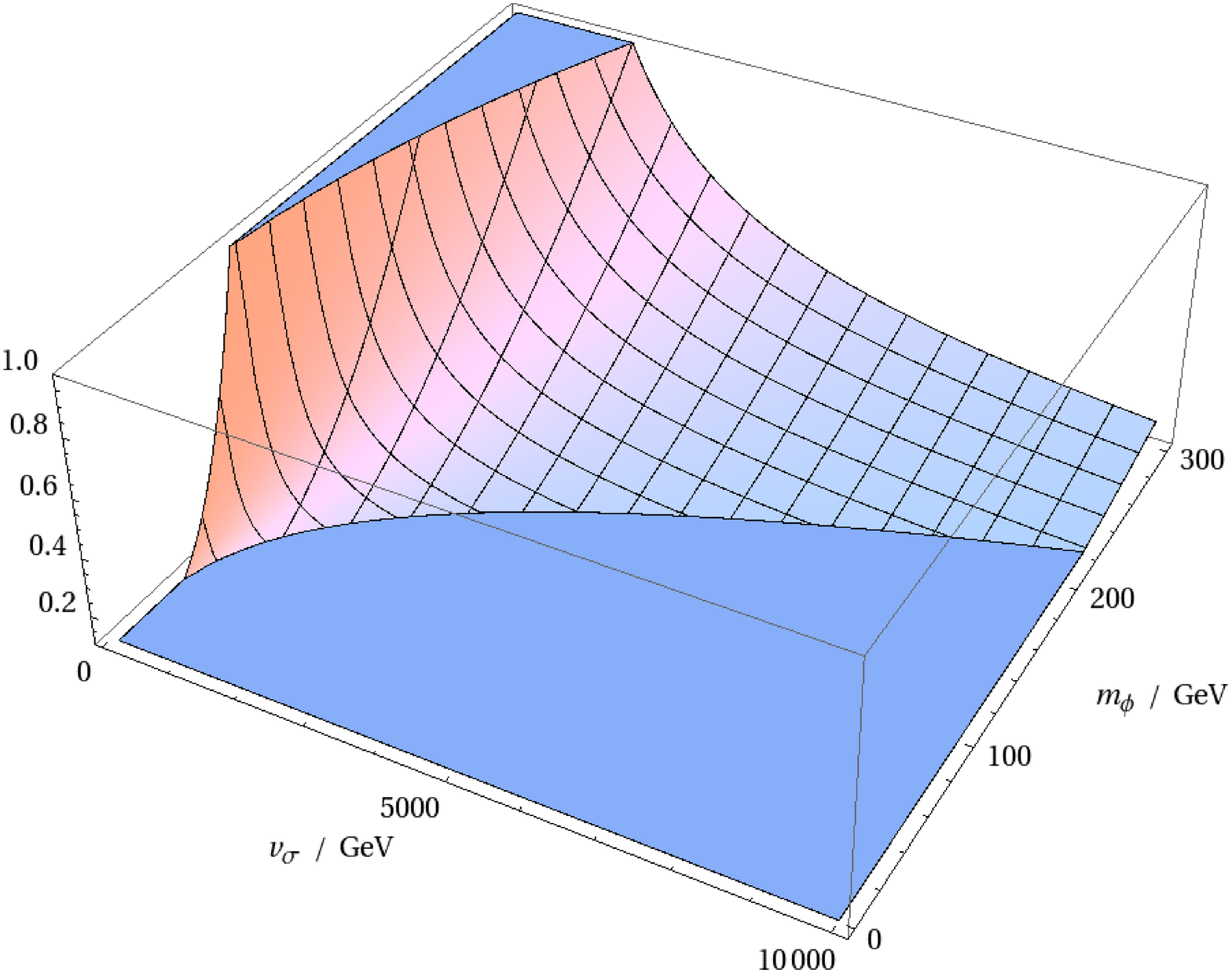,width=9cm,clip}{The $v_q$ as a function of $v_\sigma$ and $m_\phi$. \label{fig:vq} }
In an expanding Universe, before the quench, the dynamics of the rolling field are given by
\be
\ddot{\sigma}+3H\dot{\sigma}=\lambda_4\sigma^3-\lambda_6 \sigma^5.
\ee
Because of Hubble friction, energy conservation as used in (\ref{eq:sigmadot}) is in principle not strictly enforced. We can estimate the effect of the friction, say at $\sigma=\sigma_q$ by writing, for our example point 
\be
v_\sigma=3000\textrm{ GeV},\quad m_\phi=150\textrm{ GeV}\quad\rightarrow\quad \frac{3H\dot{\sigma}}{\lambda_4\sigma^3}|_{\sigma_q}=1.8\times 10^{-16},
\ee
and where we have used that $H^2\simeq V_0/3M_{\rm pl}^2$. We find in general that this ratio is less than $10^{-12}$. Hence, the force from the quartic term vastly dominates Hubble friction, which we can therefore ignore.


\section{\label{sec:triggersigma}Triggering the $\sigma$ quench}

So far, we have simply assumed that the $\sigma$ field happened to find itself at the initial value $\sigma=+\epsilon\ll1$. In fact, the initial condition is the post-inflationary equilibrium Universe at finite temperature $T$. We will now make the following additions to our model, 1) The $\sigma$ has a small negative mass term and 2) it is coupled to a (set of) light fields $\xi$. Their purpose is to provide a time-dependent mass for $\sigma$ through thermal corrections. Hence, we should add to the potential (\ref{eq:potential}),
\be
\label{eq:potential_ext}
V(\sigma,\phi)\rightarrow V(\sigma,\phi,\xi)=V(\sigma,\phi)-\frac{m_\sigma^2}{2}\sigma^2+\frac{\kappa}{2}\sigma^2\xi^2+\frac{m_\xi^2}{2}\xi^2.
\ee
 In that case, the quantum field equations read, in the Hartree approximation,
\be
\label{eq:hartree}
\ddot{\sigma}+3H\dot{\sigma}-\partial_x^2\sigma&=&-\left(-m_\sigma^2+\kappa \langle \xi^2\rangle -3\lambda_4\langle\sigma\rangle^2+15\lambda_6 \langle\sigma^2\rangle^2-g^2\langle\phi^2\rangle\right)\sigma,\\
\ddot{\phi}+3H\dot{\phi}-\partial_x^2\phi&=&-\left(m_\phi^2+3\lambda_\phi\langle\phi\rangle^2-g^2\langle\sigma^2\rangle\right)\phi,\\
\ddot{\xi}+3H\dot{\xi}-\partial_x^2\xi&=&-\left(m_\xi^2+\kappa\langle\sigma^2\rangle\right)\xi.
\ee
During this stage of the evolution, $\langle\sigma\rangle=0$, and we are in equilibrium. We are interested in the case where we wait until the temperature is very low, in which case the expansion of the Universe will eventually be dominated by the vacuum energy $V_0$. This is simply a question of waiting long enough. This also explains the choice of Hubble rate at the end of the previous section.

We also note that\footnote{Certainly for our example point $v_\sigma=3000\,$GeV, $m_\phi=150\,$GeV,} $\lambda_\phi\gg g^2 \gg\lambda_4 \simeq \lambda_6v_\sigma^2$, and we will set $\kappa\simeq \lambda_\phi$. Then we have the coupled gap equations in equilibrium
\be
\langle\phi^2\rangle = \int_k \frac{n_k^{\phi}+1/2}{\omega_k^{\phi}},\quad n_k^{\phi}=\left(e^{\omega_k^{\phi}/T}-1\right)^{-1},\\
\langle\sigma^2\rangle = \int_k \frac{n_k^{\sigma}+1/2}{\omega_k^{\sigma}},\quad n_k^{\sigma}=\left(e^{\omega_k^{\sigma}/T}-1\right)^{-1},\\
\langle\xi^2\rangle = \int_k \frac{n_k^{\xi}+1/2}{\omega_k^{\xi}},\quad n_k^{\xi}=\left(e^{\omega_k^{\xi}/T}-1\right)^{-1},
\ee
with
\be
\omega_k^{\phi}=\sqrt{m_\phi^2+3\lambda_\phi\langle\phi^2\rangle},\quad
\omega_k^{\sigma}=\sqrt{-m_\sigma^2+\kappa\langle\xi^2\rangle},\quad
\omega_k^{\xi}=\sqrt{m_\xi^2+\kappa\langle\sigma^2\rangle}.
\ee
These equations can in principle be solved numerically, although some care has to be given to renormalisation. We will not do so here, but concentrate on the important features of the system of equations.

The first equation is precisely the one we considered in the Higgs-only case in section~\ref{sec:theproblem}, although here we have not assumed a quadratic temperature dependence of the mass. In our approximation, $\phi$ decouples from the rest of the system. We will choose $m_\xi^2$ such that in the range of temperatures we are interested in $m_\xi^2\gg \kappa\langle\sigma^2\rangle$. Then the $\xi$ is just a free massive field at finite temperature providing a time-dependent mass to the $\sigma$ field. 

The middle equation (4.3) tells us, that when $\sqrt{\kappa\langle\xi^2\rangle}=m_\sigma$, the $\sigma$ field goes through a spinodal transition. This is very slow indeed, since the rate of change of the effective mass
\be
\frac{dM_\sigma^2}{dt}=\frac{d}{dt}\left(-m_\sigma^2+\kappa\langle\xi^2\rangle\right)\simeq H\kappa T^2,
\ee
is now very small (as for the case of the Standard Model Higgs in section \ref{sec:theproblem}). 

However, this does not matter, since at some point the quartic term will take over the dynamics, roughly when
\be
\sigma_{ave}=\sqrt{\langle\sigma^2\rangle}\simeq \sqrt{\frac{m_\sigma^2}{\lambda_4}}.
\ee
We only need to make sure that this is well before the $\phi$ quench, i.e. choose $m_\sigma$ such that,
\be
\sqrt{\frac{m_\sigma^2}{\lambda_4}}\ll \frac{m_\phi}{g},
\ee
which in our example amounts to $m_\sigma\ll 15\,$GeV. This choice also means that the location of $v_\sigma$ (\ref{eq:vevs}) and the estimate of $\dot{\sigma}$ (\ref{eq:sigmadot}) is unaffected by the introduction of $m_\sigma$. Finally, choosing $m_\sigma$ small enough ensures that the temperature of the Universe is well below $100\,$GeV, as required by Cold Electroweak Baryogenesis.

In principle, all IR modes with $|k|<m_\sigma$ will participate in the spinodal ``roll-off''. For our purpose, we will simply let the quantity $\sigma_{ave}$ play the role of $\sigma$, when considering the subsequent dynamics as in section \ref{sec:triggerhiggs}. 

Again, we need to make sure that the dynamics will not be Hubble friction dominated, and we write
\be
v_\sigma=3000\textrm{ GeV},\quad m_\phi=150\textrm{ GeV}\quad\rightarrow \quad\frac{3H\dot{\sigma}}{m_\sigma^2\sigma}|_{\sigma_q}=4\times 10^{-14},\quad m_\sigma=1\textrm{ GeV}.
\ee
Note that although a spinodal transition is triggered the moment $M^2_\sigma<0$, this transition is {\it very} slow at first, and involves only the very IR modes\footnote{In the approximation $\langle\xi^2\rangle\propto T^2$, one may solve the field evolution exactly by writing $M_\sigma^2(t)=m_\sigma^2(e^{-2Ht}-1)$ \cite{Aarts:2007qu} or in the linear approximation $M_\sigma^2(t)=-m_\sigma^2(2Ht)$ \cite{GarciaBellido:2002aj}. }. Asymptotically, of course, $M^2_\sigma\rightarrow -m_\sigma^2$, and the transition will complete, and even before that the quartic term in the potential will have taken over. 


\section{\label{sec:checks}Defects, bubbles and loops}

\EPSFIGURE[t]{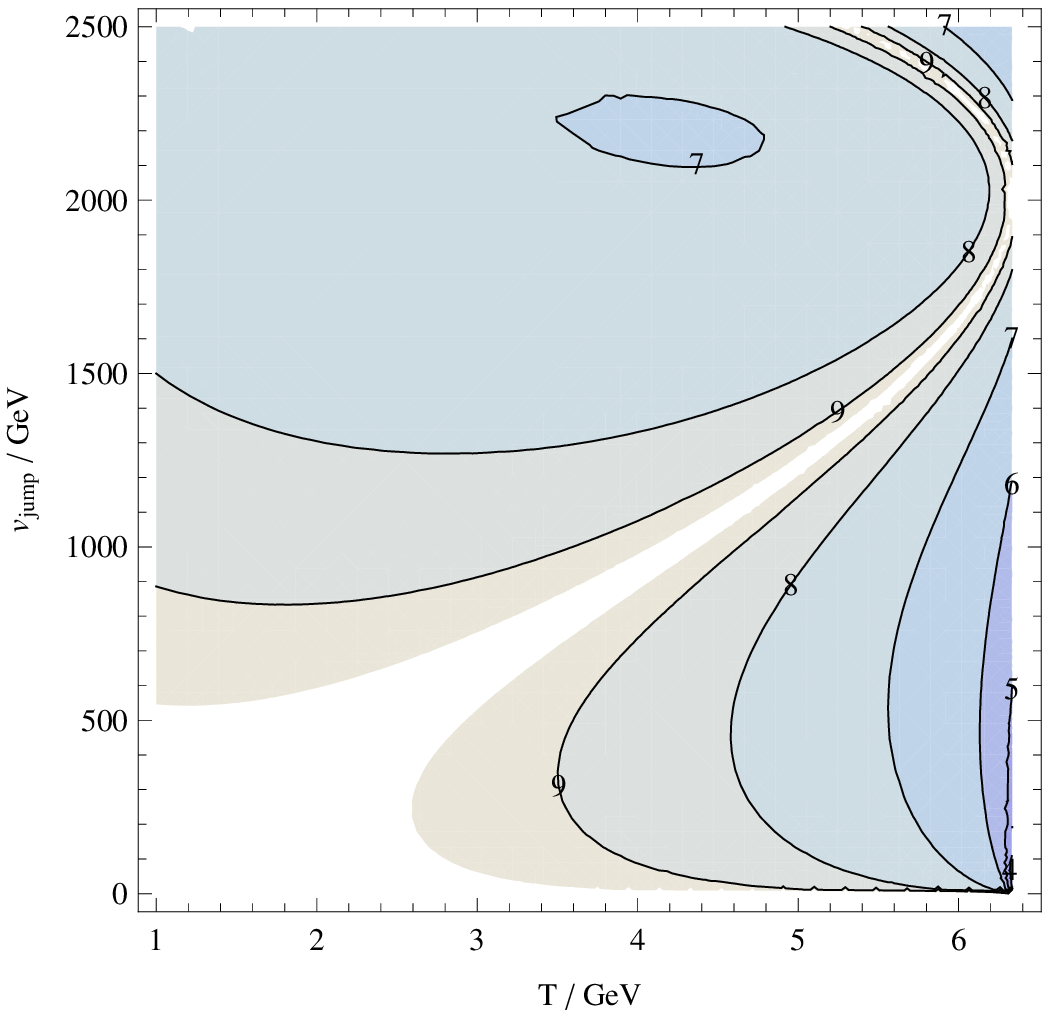,width=9cm,clip}{Contour plot of the energy of the critical bubble in units of the temperature, $\log_{10}(E_{\rm crit}/T)$, for our example point $v_\sigma=3000\textrm{ GeV},\quad m_\phi=150\,$GeV. \label{fig:nucrate} }

We first note that the $\sigma$ potential has $Z_2$ symmetry, and topological defects (kinks) potentially form in the transition. However, because the transition is so slow at first, the density of kinks will be very small. In fact, the dynamics may be so slow that even before $M^2_\sigma<0$, a first order phase transition may be triggered, with nucleation of bubbles. In particular, a first order phase transition is possible at finite temperature, even though the tree-level/zero-temperature potential does not have a ``bump''. Whether this happens depends on the details of the potential, and an accurate determination of the bubble nucleation rate requires careful numerical simulations \cite{Moore:2001vf}.

However, a simple estimate of the nucleation rate can be given in terms of the critical bubble
\be
\frac{\Gamma_{\rm Nucleation}(T)}{\mathcal{V}}\simeq T^4e^{-\frac{E_{\rm crit}}{T}},\qquad E_{\rm crit}=\frac{16\pi}{3}\frac{\Sigma(v_{\rm jump})^2}{\Delta V^3},
\ee
where $E_{\rm crit}$ is the energy of the critical bubble, $v_{\rm jump}$ is the field expectation value that the field jumps to inside the bubble, the ``latent heat'' is $\Delta V=V(\sigma=0)-V(\sigma=v_{\rm jump})$ and the wall tension is given by
\be
\Sigma(v_{\rm jump})=\int_0^{v_{\rm jump}}\sqrt{2(V(v)-V(v_{\rm final}))}\,dv.
\ee
Let us first ignore the Higgs
and consider the effective potential
\be
\tilde{V}(\sigma)=\tilde{V_0}+\frac{\tilde{\kappa} T^2- m_\sigma^2}{2}\sigma^2-\frac{\lambda_4}{4}\sigma^4+\frac{\lambda_6}{6}\sigma^6.
\ee
with the minimum $v_{\rm final}$, and where $\tilde{V_0}$ is fixed so that $\tilde{V}(v_{\rm final})=0$. One can find $v_{\rm jump}$ by the over-shoot/under-shoot method, but we will simply calculate the exponent $\frac{E_{\rm crit}}{T}$, as a function of $T$ and $v_{\rm jump}$, for $\kappa=1$. We need to compare the nucleation rate within a volume $\mathcal{V}\simeq H^{-3}$ to the Hubble rate $H$. Since $\sigma=0$ becomes the global minimum for $T> 7\,$GeV, negligible nucleation amounts to
\be
\label{eq:tunconstraint}
\frac{E_{\rm crit}}{T}\gg \ln\left(\frac{T^4}{H^4}\right)\simeq 130.
\ee
Using again our example point, Fig.~\ref{fig:nucrate} shows the contours of $\log_{10}(E_{\rm crit}/T)$. In fact, the lower right triangle corresponds to ``tunneling'' that does not pass the bump, and so corresponds to nucleation of a subleading bubble which would contract again. 

When including the Higgs, we should consider the option of tunneling directly into the global minimum $(v_\sigma,v_\phi)=(3000\,\textrm{GeV},246\,\textrm{GeV})$. To estimate the impact on the rate, we can compare the potential at the global minimum to the potential at the maximum tunneling rate region from Fig.~\ref{fig:nucrate}, around $(\sigma,\phi)=(2250\,\textrm{GeV},0\,\textrm{GeV})$,
\be
\frac{V(0,0)-V(2250,0)}{V(0,0)-V(3000,246)}\simeq 0.91.
\ee
Hence, there is no dramatic gain in ``latent heat'' from tunneling directly to the global minimum, to compensate for the increase in wall tension from tunneling further in field space. We conclude that the constraint (\ref{eq:tunconstraint}) is easily met, and we therefore do not expect bubble nucleation to play a role in this model. \footnote{ This turns out to be a result of having a rather small $\lambda_4$ and correspondingly a very large $v_{\rm jump}$, compared to for instance a similar calculation in the Standard Model Higgs potential (for Higgs masses where there is a first order phase transition).} 

Finally, we emphasize that we consider (\ref{eq:potential}), (\ref{eq:potential_ext}) to be an effective potential, and in that sense, quantum corrections are already included. However, as a check of consistency, it is relevant to consider corrections to the $\sigma$ mass at the one-loop level, since our scenario is based on $m_\sigma^2$ being negative (and small). Following \cite{vanTent:2004rc}, the contribution to the effective potential is simply
\be
\label{eq:loop}
V^{1}=\sum_i\frac{(m_i^2(\sigma))^2}{64\pi^2}\ln\left(\frac{m_i^2(\sigma)}{\nu^2}\right),
\ee
where $\nu$ is a renormalisation scale, and $m_i^2(\sigma)$ are the eigenvalues of the mass matrix (including $\xi$), at a given value of $\sigma$ (and $\phi$ and $\xi$). Away from $\sigma=0$, we need to include (\ref{eq:loop}) for all $m_i$, and we run into the well-known problem of negative eigenvalues (for small $\sigma$ in the $\sigma$ direction; and near $\sigma_q$ in the $\phi$ direction), making the effective action complex\footnote{The effective action can be calculated by other methods such as a 2PI resummation or full-fledged lattice Monte-Carlo, or even by invoking some Maxwell construction, but this is far beyond the scope of the present work.}. Given the correct effective potential, we could solve for a different set of ``bare'' parameters to satisfy the original constraints. But at one loop (\ref{eq:loop}), the Higgs quench rate constraint is beyond our reach.
 
We will therefore restrict ourselves to considering the $\sigma$ mass term around $\sigma=\phi=\xi=0$, and since $\kappa\gg g^2\simeq\lambda_4$, we concentrate on the contribution from the $\xi$ field (which is real). We use $m_i^2=m_\xi^2+\kappa\sigma^2$ to find
\be
\left(\frac{d^2V^1}{d\sigma^2}\right)_{\sigma=\phi=\xi=0}=\frac{\kappa m_\xi^2}{32\pi^2}\left(1+2\ln\left(\frac{m_\xi^2}{\nu^2}\right)\right).
\ee
We will assume a renormalisation scale $\nu>m_\xi$, so that the log is negative. But even ignoring the log, as long as $m_\xi<\sqrt{32}\pi/\kappa |m_\sigma|$, the effective potential still has negative curvature at $\sigma=0$.
\footnote{The constraint on $m_\xi$ introduced above, $m_\xi\gg \sqrt{\kappa\langle\sigma^2\rangle}\simeq 1\,$GeV, was only for practical purposes, and is not essential.}


\section{\label{sec:electroweak}Electroweak physics}

\EPSFIGURE[t]{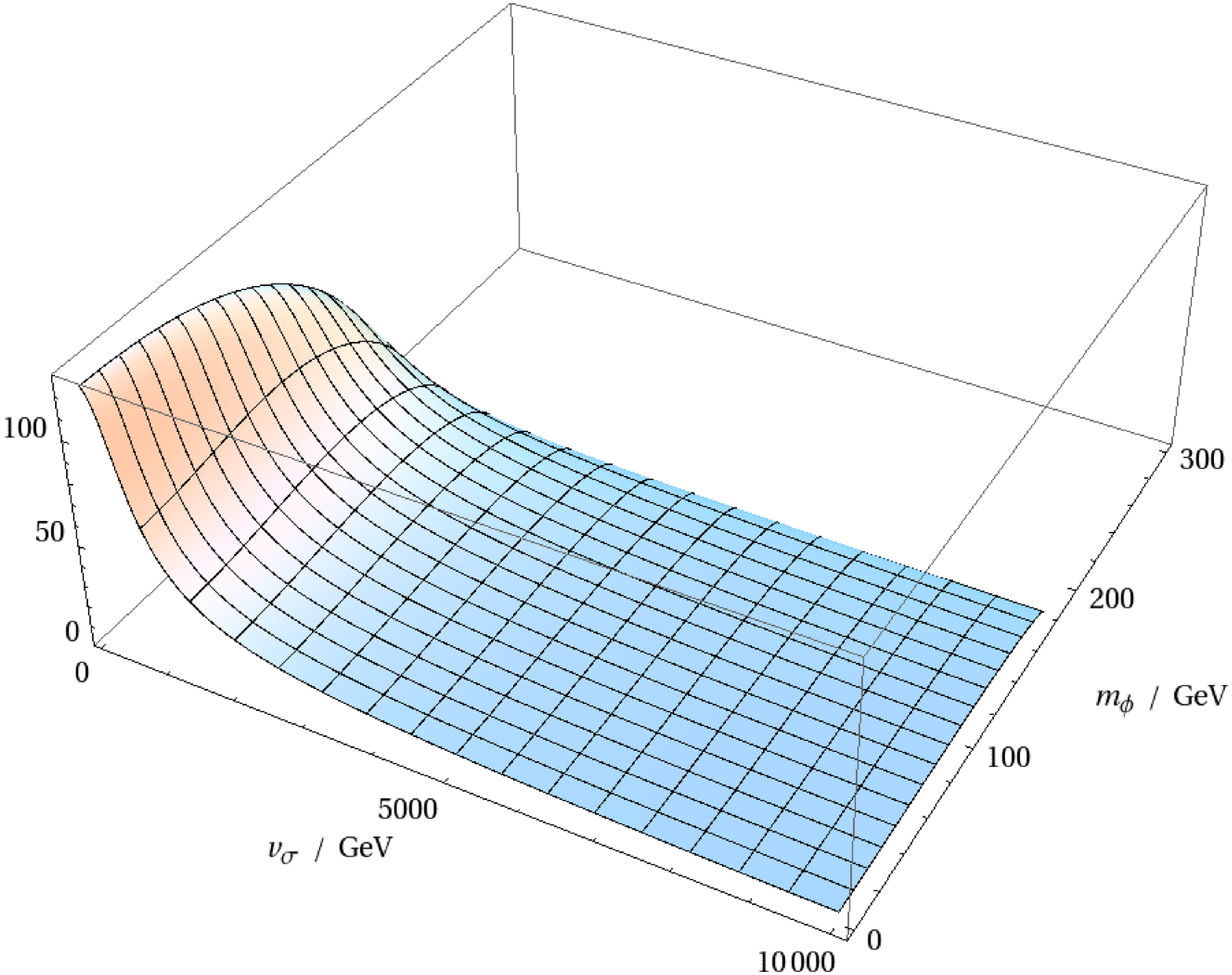,width=9cm,clip}{The smaller mass eigenvalue $m_-$ as a function of $v_\sigma$ and $m_\phi$. \label{fig:m2} }
In order not to conflict with known electroweak physics, we fixed the $\phi$ vacuum expectation value to $v_\phi=246\,$GeV. Because of the $\phi-\sigma$ coupling, there is mixing in the vacuum, with a mass matrix that reads
\be
M^2=
\left(
\begin{array}{cc}
2\lambda_\phi v_\phi^2
&
-2g^2v_\sigma v_\phi
\\
-2 g^2 v_\sigma v_\phi
&
-2\lambda_4v_\sigma^2+4\lambda_6v_\sigma^4.
\end{array}
\right).
\ee
This can be diagonalised in a straightforward way, and Fig.~\ref{fig:m2} shows $m_-$ as a function of $v_\sigma-m_\phi$, corresponding to the eigenvalues $_\pm$ of the mass matrix. We see that for $m_\phi>182\,$GeV, the vacuum becomes unstable, ruling out that part of parameter space. For our example point, we have
\be
v_\sigma=3000\textrm{ GeV},\quad m_\phi=150\textrm{ GeV}\quad\rightarrow\quad m_+=164\textrm{ GeV},\quad m_-=18\textrm{ GeV},
\ee
with a mixing angle
\be
\theta=\tan^{-1}\left(\frac{m_+^2-2\lambda_\phi v_\phi^2}{2g^2v_\sigma v_\phi}\right)=0.225,\qquad \theta=13^{\circ}.
\ee

One may worry that when the mostly-$\phi$  mode $m_+$ is heavier than the mostly-$\sigma$ mode $m_-$, reheating will proceed into mostly-$\sigma$ particles rather than the Standard Model degrees of freedom. However, because of the smallness of $g^2$, this does not happen. For instance, the $\phi\rightarrow 2\sigma$ decay mediated by the term\footnote{Expanding around the vacuum, $\sigma=v_\sigma+\delta\sigma$, $\phi=v_\phi+\delta\phi$} $2g^2v_\phi\delta\phi\delta\sigma^2$ , is 
\be
\Gamma=\frac{g^4v_\phi^2}{8\pi m_H}\sqrt{1-\frac{4M_-^2}{M_+^2}}\simeq 2.26\times 10^{-4}\,\textrm{GeV}.
\ee
Comparing this to the total width of Higgs decay into Standard Model degrees of freedom in this mass range, $\Gamma\simeq (10^{-3}-1)\,$GeV \cite{Eidelman:2004wy}, we conclude that the vast majority of the available energy will be channeled into these.

\EPSFIGURE{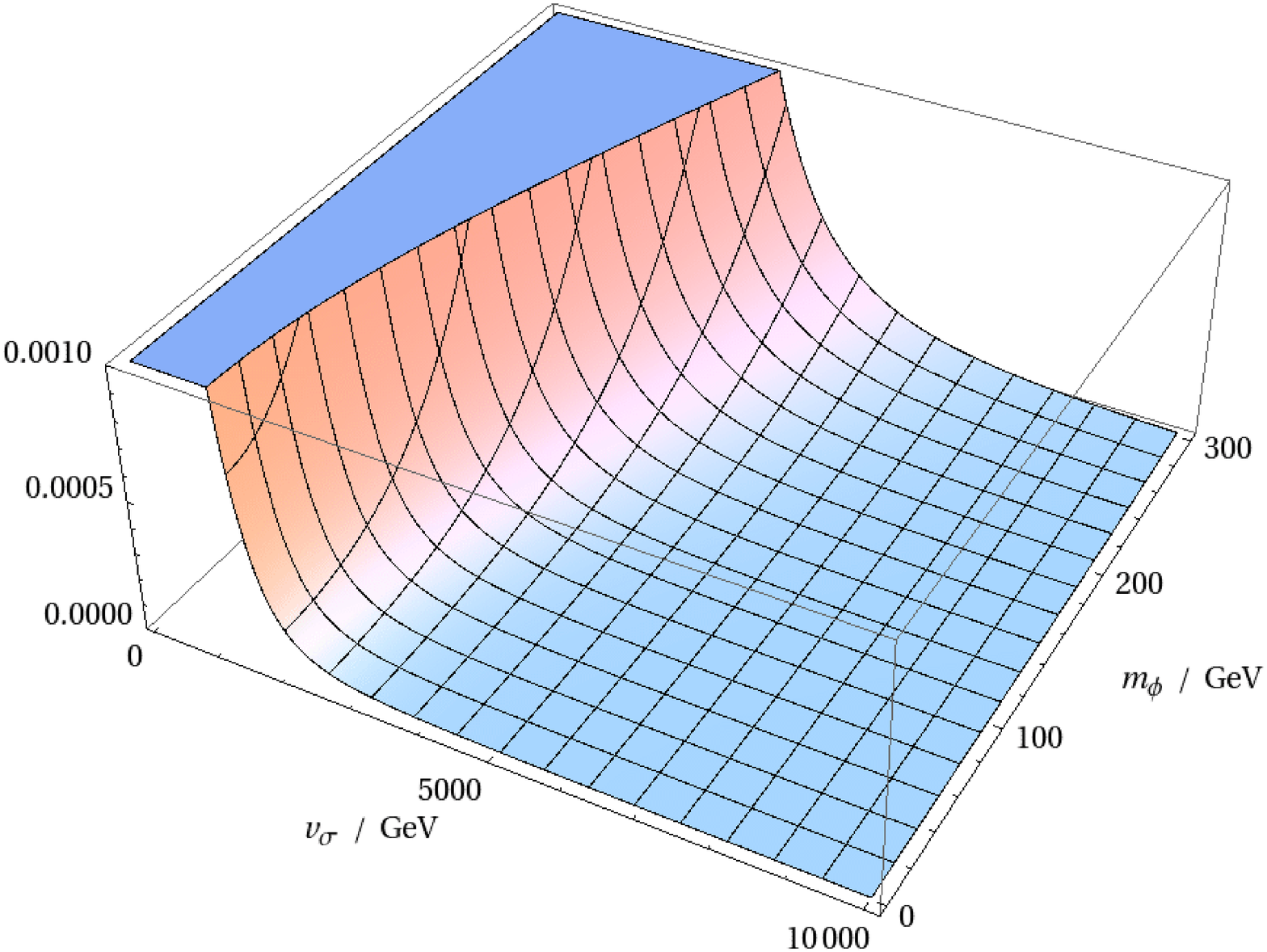,width=9cm,clip}{The mostly-$\phi$ to two mostly-$\sigma$ decay rate $\Gamma$ as a function of $v_\sigma$ and $m_\phi$. \label{fig:gamma} }


\section{\label{sec:theconclusion}Conclusion}

Using a simple implementation of the Higgs-$\sigma$ potential, we have argued that the expansion of the
Universe can be responsible for triggering a fast electroweak symmetry breaking transition. At the same time,
this transition can be delayed until the temperature of the Universe is far below the electroweak scale.

The model can be summarised in the potential
\be
V(\sigma,\phi)=V_0-\frac{m_\sigma^2}{2}\sigma^2-\frac{\lambda_4}{4}\sigma^4+\frac{\lambda_6}{6}\sigma^6-\frac{g^2}{2}\sigma^2\phi^2+\frac{m_\phi^2}{2}\phi^2+\frac{\lambda_\phi}{4}\phi^4+\frac{\kappa}{2}\sigma^2\xi^2+\frac{m_\xi^2}{2}\xi^2,\nonumber\\
\ee
and we split the evolution into 3 stages: 
\begin{itemize}
\item 1) The post inflationary Universe in equilibrium cools under Hubble expansion until the $\sigma$ is driven to a tachyonic transition.
\item 2) When $\sigma_{\rm ave}=\langle\sigma^2\rangle\simeq m_\sigma^2/\lambda_4$ Hubble expansion can ignored and $\sigma$ rolls down its potential, triggering a fast tachyonic quench of the Standard Model Higgs, $\phi$.
\item 3) $\phi$ and $\sigma$ settle in the global minimum of the potential, for which the physics is constrained by electroweak phenomenology.
\end{itemize}

\EPSFIGURE[t]{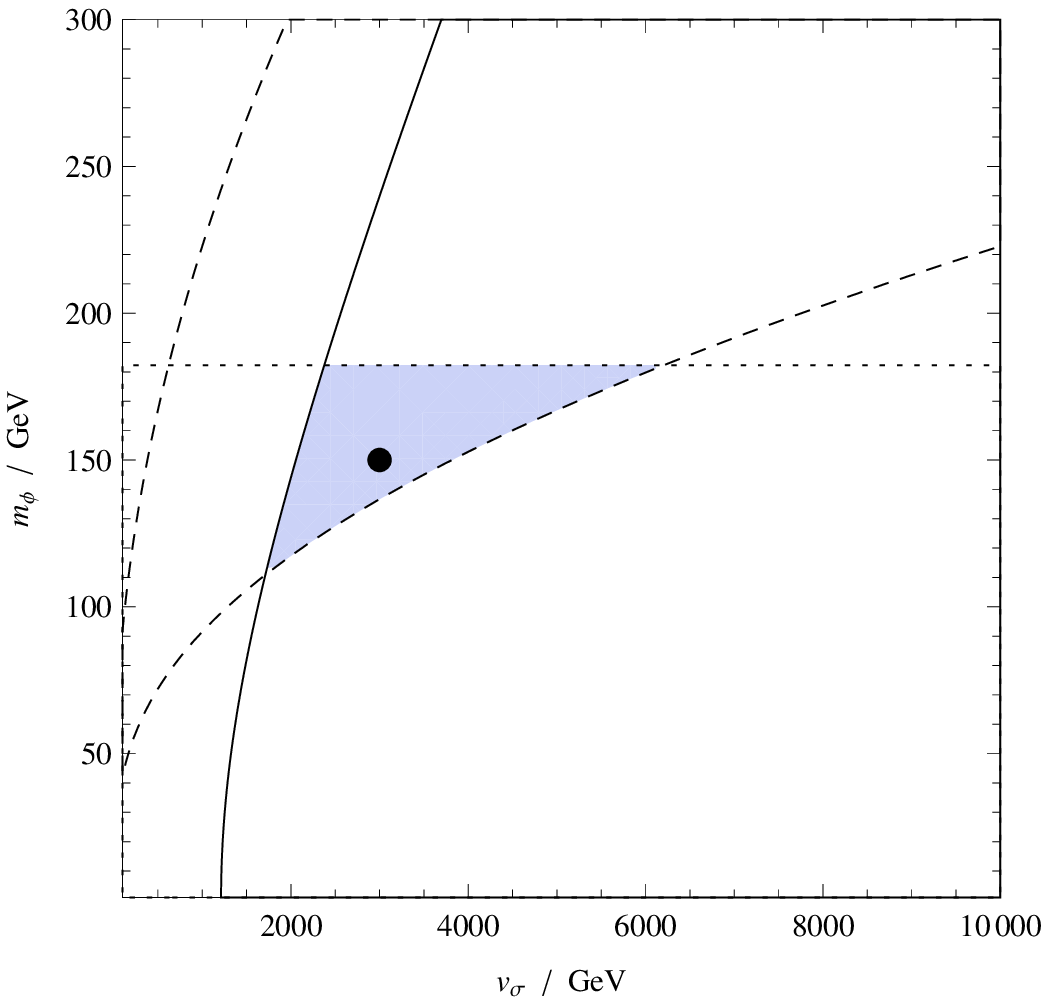,width=9cm,clip}{The region in parameter space consistent with all three constraints presented in Figs.~\ref{fig:vq}, \ref{fig:m2} and \ref{fig:gamma}. The dot is our example point $v_\sigma=3000\,$GeV, $m_\phi=150\,$GeV. \label{fig:final} }

The cost is the addition of an extra scalar field coupling to some other matter sector ``$\xi$'', and with self-couplings $\sigma^4$ and $\sigma^{6}$. This is partly a result of insisting that electroweak physics is unaltered, but it also
comes from restricting ourselves to a minimal model. More complicated $\sigma$-potentials can be envisaged, and for instance replacing ($\sigma^4$, $\sigma^6$) by ($\sigma^5$, $\sigma^6$) or ($\sigma^6$, $\sigma^8$) works as well. The input values used here for the Higgs mass ($m_H=160\,$GeV) can be relaxed within the range still allowed by experiment, say $m_H\in[115:200]\,$GeV. Similarly, the height of the potential $V_0$ can be relaxed to $\simeq 200^4\,$GeV, while still avoiding finite temperature electroweak symmetry restoration after reheating. 

The model is very similar to the low-scale inflation model proposed in \cite{Copeland:2001qw,vanTent:2004rc},
although since now $\sigma$ is not required to be responsible for inflation and the CMB fluctuations, a simpler
potential and weaker tuning is allowed. Still, as in \cite{vanTent:2004rc} there is $\phi$-$\sigma$ coupling
and mixing, providing an observational signature of this model. However, in contrast to \cite{vanTent:2004rc}, one mass eigenvalue ($m_-$) is smaller than $m_H$($\simeq m_+$), and the Higgs can in principle decay into $\sigma$ particles. This is however suppressed by a small $\phi-\sigma$ coupling.

We conclude that although the expansion of the Universe is very slow at the electroweak scale, it can be amplified in a rather straightforward way to provide a fast electroweak transition. This transition can be delayed indefinitely (essentially until $T\simeq m_\sigma$), and so also until the Universe is cold enough that the computation \cite{Hernandez:2008db} of the effective Standard Model CP-violation becomes reliable. And so, although far from being the final word on the matter, we consider the scenario presented here an interesting option to realise a cold, fast electroweak transition.

\acknowledgments
We thank Jan Smit for useful comments. This research is supported by the
European Union through Marie Curie Research and Training Network
``UNIVERSENET'' (MRTN-CT-2006-035863). K.E. is also supported by the Academy of Finland
grants 218322 and 131454. O.T. is supported by the Magnus Ehrnrooth Foundation.

\end{document}